\documentclass[aps,prd,amssymb,amsmath,onecolumn,preprint,superscriptaddress]{revtex4-1}

\usepackage{graphicx}
\usepackage{epstopdf}
\usepackage{amsmath}
\usepackage[section]{placeins}
\usepackage{slashed}
\usepackage{color}
\usepackage{soul}
\usepackage{appendix}
\usepackage{placeins}
\usepackage[utf8]{inputenc}
\usepackage{pstricks}
\usepackage{multirow}
\usepackage{hyperref}
\usepackage{subfigure}
\usepackage{epsfig}
\usepackage{cancel}
\usepackage{braket}
\usepackage[normalem]{ulem} 
\newcommand{\be}{\begin{equation}}
\newcommand{\ee}{\end{equation}}
\newcommand{\ben}{\begin{eqnarray}}
\newcommand{\een}{\end{eqnarray}}

\usepackage{forest}
\usepackage{xcolor}
\usepackage{amssymb}
\usepackage{txfonts}
\usepackage{feynmp-auto}
\usepackage{simpler-wick}
\usetikzlibrary{shapes,arrows,positioning,automata,backgrounds,calc,er,patterns}
\usepackage[left=2.5cm,top=2.5cm,right=2.5cm,bottom=2.5cm]{geometry}
\usepackage[compat=1.1.0]{tikz-feynman}
\usepackage{tikz}
\usepackage{pgfkeys}
\usetikzlibrary{tikzmark}
\usepackage{stackengine}
\graphicspath{{Figuras/}}
\usepackage{subcaption}

\begin{document}

%
%


\title{Final-state interactions in the $B^{\mp}\to K^+K^- \pi^{\mp} $ decays: traces of the $X(3700)$ state and CP asymmetry}

\author{Luciano M. Abreu}
\email{luciano.abreu@ufba.br}
\affiliation{Instituto de F\'isica, Universidade Federal da Bahia, Campus Universit\'ario de Ondina, 40170-115, Bahia, Brazil}
\author{Eduardo Almeida}
\email{almeidae@ufba.br}
\affiliation{Instituto de F\'isica, Universidade Federal da Bahia, Campus Universit\'ario de Ondina, 40170-115, Bahia, Brazil}
\author{Patrícia C. Magalhães}
\email{pcmaga@unicamp.br}
\affiliation{Instituto de F\'isica Gleb Wataghin, Universidade de Campinas, Brazil}
\author{Vin\'icius S. Silva}
\thanks{Corresponding author}
\email{viniciussouza@ufba.br}
\affiliation{Instituto de F\'isica, Universidade Federal da Bahia, Campus Universit\'ario de Ondina, 40170-115, Bahia, Brazil}
\begin{abstract}

In this work, we investigate final-state interaction effects in $B^{\mp}\to K^+K^- \pi^{\mp}$ decays above 3 GeV with a twofold objective: (i) to study $CP$ asymmetry, and (ii) to search for possible signatures of the $X(3700)$ state (also denoted as $X(3720)$). We consider both tree-level and resonant contributions, with the latter arising from a triangle mechanism $B^{\mp} \to \bar{D}^0 D^{\ast \mp} (\to \pi^{\mp} D^0) \to K^+K^- \pi^{\mp}$, in which the $D \bar D \to K \bar K$ rescattering process carries the signature of the exotic $X(3700)$ state. We analyze the dependence of our results on the relative phase between the tree-level and resonant contributions, as well as on other relevant parameters. We estimate the differential mass distribution in the region where the $X(3700)$ may leave a trace, and the $CP$ asymmetry driven by final-state interactions. Our findings are compared with available data for the differential mass distributions and the $CP$ asymmetry. Taken together, these results may contribute to a better understanding of the properties of exotic states and the role of $CP$ asymmetry in $B$-meson phenomenology.

\end{abstract}

\maketitle

\section{Introduction}

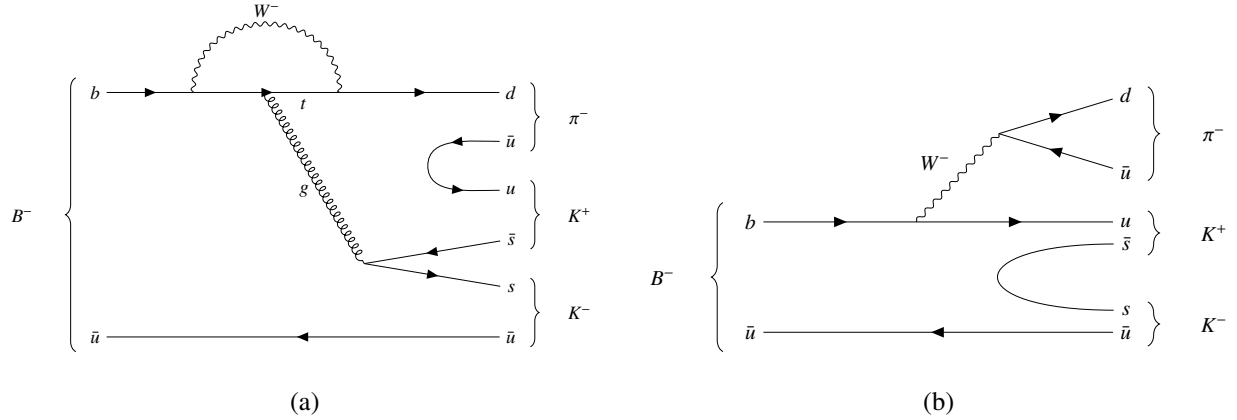
\begin{figure}[htbp]
\subfigure[]{
    \centering
    \resizebox{0.48\textwidth}{!}{
      \begin{tikzpicture}
    \begin{feynman}
      \vertex (b0) at (0, 2.5) {\(b\)};
      \vertex (u0) at (0, -2.5) {\(\bar{u}\)};

      \vertex (v1) at (2.0, 2.5);      
      \vertex (v2) at (3.5, 2.5);      
      \vertex (v3) at (5.0, 2.5);      

      \vertex (d) at (8.5, 2.5) {\(d\)};
      \vertex (u_spec) at (8.5, -2.5) {\(\bar{u}\)};

      \vertex (vac_ubar) at (8.5, 1.5) {\(\bar{u}\)};
      \vertex (vac_u)    at (8.5, 0.5) {\(u\)};
      \vertex (vac)      at (6.8, 1.0); 

      \vertex (g_split) at (5.5, -1.0);
      \vertex (sbar)    at (8.5, -0.5) {\(\bar{s}\)};
      \vertex (s)       at (8.5, -1.5) {\(s\)};

      \diagram* {
        (b0) -- [fermion] (v1) -- [fermion] (v3) -- [fermion] (d),
        
        (v1) -- [boson, half left, looseness=1.6, edge label=\(W^-\)] (v3),
        
        (v2) -- [gluon, edge label'=\(g\)] (g_split),
        
        (sbar) -- [fermion] (g_split) -- [fermion] (s),

        (vac_ubar) -- [fermion, out=180, in=90] (vac) -- [fermion, out=-90, in=180] (vac_u),

        (u_spec) -- [fermion] (u0)
      };

      \node[below] at (4.25, 2.5) {\(t\)};

      
      \draw[decorate, decoration={brace, mirror, amplitude=6pt}]
        (-0.4, 2.8) -- (-0.4, -2.8)
        node[midway, xshift=-1.1cm] {\(B^-\)};

      \draw[decorate, decoration={brace, amplitude=5pt}]
        (8.9, 2.7) -- (8.9, 1.3)
        node[midway, xshift=1.0cm] {\(\pi^-\)};

      \draw[decorate, decoration={brace, amplitude=5pt}]
        (8.9, 0.7) -- (8.9, -0.7)
        node[midway, xshift=1.0cm] {\(K^+\)};

      \draw[decorate, decoration={brace, amplitude=5pt}]
        (8.9, -1.3) -- (8.9, -2.7)
        node[midway, xshift=1.0cm] {\(K^-\)};

    \end{feynman}
  \end{tikzpicture}
    }
         \label{fig1a}
    }
  \hfill
\subfigure[]{
    \centering
    \resizebox{0.48\textwidth}{!}{
\begin{tikzpicture}
\begin{feynman}
\vertex (b0) at (0,1) {\(b\)};
\vertex (u0) at (0,-1) {\(\bar u\)};
\vertex (v1) at (3,1);
\vertex (u1) at (6.8,1) {\(u \)};
\vertex (u2) at (6.8,-1) {\(\bar u \)};
\vertex (w) at (4.5,2.6);
\vertex (d) at (6.8,3.3) {\(d\)};
\vertex (ubar) at (6.8,1.9) {\(\bar u\)};
\vertex (sbar) at (6.8,0.6) {\(\bar s \)};
\vertex (s) at (6.8,-0.6) {\(s \)};
\diagram*{
(b0) -- [fermion] (v1) -- [fermion] (u1),
(u2) -- [fermion] (u0),
(v1) -- [boson, edge label=\(W^-\)] (w),
(w) -- [fermion] (d),
(ubar) -- [fermion] (w),
(sbar) -- [half right, looseness=6] (s)
};
\draw[decorate,decoration={brace,mirror,amplitude=6pt}]
(-0.5,1.35) -- (-0.5,-1.35)
node[midway,xshift=-1.1cm] {\(B^-\)};
\draw[decorate,decoration={brace,amplitude=6pt}]
(7.2,3.45) -- (7.2,1.75)
node[midway,xshift=1.2cm] {\(\pi^-\)};
\draw[decorate,decoration={brace,amplitude=6pt}]
(7.2,1.2) -- (7.2,0.4)
node[midway,xshift=1.2cm] {\(K^+ \)};
\draw[decorate,decoration={brace,amplitude=6pt}]
(7.2,-0.45) -- (7.2,-1.25)
node[midway,xshift=1.2cm] {\(K^-  \)};
\end{feynman}
\end{tikzpicture}
    }
         \label{fig1b}
}
\caption{Topologies contributing to the
  $B^{\mp}\to K^+K^-\pi^{\mp}$ decay: [\ref{fig1a}] penguin diagram;  
 and [\ref{fig1b}] $W^{-}$ external-emission  $b\to u $  
     tree-level  diagram. }
  \label{fig:weakTopologies}
\end{figure}

$B$ meson decays offer a rich laboratory for studying $CP$ violation and the spectroscopy of hadronic states. On the $CP$ violation side, experimental results have shown that $B^{\mp}\to h^+h^- h^{\prime \mp}$ ($h = K, \pi$) decays display a rich distribution of $CP$ asymmetry $(A_{CP})$  ~\cite{LHCb:2013lcl,LHCb:2014mir,LHCb:2019xmb,LHCb:2019jta,LHCb:2019sus,LHCb:2022fpg}. The generation of $CP$ asymmetries requires the interference of different strong and CKM weak phases. This is standardly achieved through the interference between penguin and tree Feynman diagrams, as depicted in Fig.~\ref{fig:weakTopologies} for the $B^{\mp}\to K^+K^- \pi^{\mp}$ decay considered in this work.

However, this alone is not sufficient to explain the data ( see Ref. \cite{Bediaga:2020qxg} for an extensive  review). 
Alternatively, the observed $CP$ phenomenon may be explained by the interference between different sources, namely weak topologies, and hadronic final state interactions, such as $\pi^+\pi^- \to K^+ K^-$ or $D^0 \bar D^0 \to K^+ K^-$, that mix states through coupled-channel effects~\cite{Bediaga:2017axw,Bediaga:2020ztp,magalhaes2023,Heuser:2025mnk}. 

A recently amplitude analysis of the $B^{\mp}\to K^+K^- \pi^{\mp}$ decay \cite{LHCb:2019xmb} has shown that the $K\pi$ resonances produced from the penguin diagram are suppressed in respect to $ K \bar K$ resonances with isospin zero (known to be strongly coupled to $\pi\pi$ channel) produced from the tree diagram. It was also observed that $K \bar K$ $(s\bar s)$ resonances with isospin 1 can only be produced through a OZI-suppressed diagram \cite{Gronau_2008}. 
The relevance of the hadronic final state interaction in this decay is evident in the context of the $CP$ violation. LHCb reported in \cite{LHCb:2019xmb} that the $\pi\pi \to KK$ rescattering contributes  $16\%$ to the branching-fraction, with $66\%$ of $CP $ asymmetry.
 
Given this scenario, in this work we are interested on the $A_{CP}$ that also arises in the high-mass region of the phase space, as can be seen, for example, in the data of $B^{\mp}\to K^+K^- \pi^{\mp}$ decays reported by the LHCb Collaboration~\cite{LHCb:2022fpg}. To the best of our knowledge, the specific source of $A_{CP}$ in the high-mass region of this channel has not yet been identified. One may therefore wonder whether it originates from  a strong-phase variation across the phase space, as was proposed  in similar decays analyzed in Refs.~\cite{Bediaga:2017axw,Bediaga:2020ztp}. In this case, such a strong-phase variation could stem from final-state interactions mediated by $D^0 \bar D^0 \to K^+ K^-$ rescattering sourced by another weak topology, Fig.~\ref{d1}, with a triangle loop involving charmed mesons that produce the light-meson final state. The loop integral naturally gives rise to a strong phase that varies with the $K^+ K^-$ invariant mass. 

On the exotic hadron spectroscopy side, the nature of the $X(3700)/X(3720)$ state (hereinafter denoted as $X(3700)$) remains an open question. It has been interpreted as the lightest partner of the $X(3872)$, i.e., a possible $D\bar{D}$ bound state with $I^G(J^{PC}) = 0^+(0^{++})$, arising from the strong interaction between charmed mesons~\cite{Gamermann:2006nm,Gamermann:2007mu,Wang:2019evy,Ding:2020dio,Xiao:2012iq,Brandao:2023vyg,Dai:2020yfu,Sobrinho:2024tre,Abreu:2025mhl,Wang:2020elp,Deineka:2021aeu}. Experimental searches reported in the literature are scarce. The Belle and BaBar Collaborations analyzed the $e^+ e^- \to J/\psi D \bar{D}$ and $e^+ e^- \to D \bar{D}$ reactions, respectively, but found no clear identification~\cite{Belle:2007woe,Belle:2017egg,BaBar:2010jfn}. On the other hand, some authors claim that these data support the observation of the $X(3700)$~\cite{Gamermann:2007mu,Xiao:2012iq,Wang:2020elp,Deineka:2021aeu}. Nevertheless, this structure has not yet been listed in the summary tables of the Review of Particle Physics~\cite{ParticleDataGroup:2024cfk}.

In a recent study using the local hidden-gauge formalism, it was proposed that the $X(3700)$ could leave detectable signatures in the $\eta\eta$ invariant mass distribution of the $B^{+}\to K^+ \eta \eta$ reaction~\cite{Brandao:2023vyg}, due to the strong coupling of the $\eta\eta$ system to the $X(3700)$. These findings motivate our present investigation, in which we explore how the $X(3700)$ manifests in the rescattering processes of $B^{\mp}\to K^+K^- \pi^{\mp}$ decays.

Thus, stimulated by the discussions above, in this work we investigate final-state interaction effects in $B^{\mp}\to K^+K^- \pi^{\mp} $ decays with two objectives: the study of $CP$ asymmetry and the search for possible signatures of the $X(3700)$. We focus on the massa region above 3 GeV, considering both tree-level and resonant contributions. The former sources the $KK$ and $K\pi$ low energy resonances, which above 3 GeV is dominated by tail. The latter arises from a triangle mechanism $B^{\mp} \to \bar{D}^0 D^{\ast \mp} (\to \pi^{\mp} D^0) \to K^+K^- \pi^{\mp}$, generating the final-state interaction via the $D \bar D \to  K \bar K$ rescattering process, which carries the signature of the exotic $X(3700)$. 
We analyze the dependence of our results on the relative phase between the tree-level and resonant contributions, as well as on other relevant parameters. The findings are also compared with available data for the differential mass distributions and the $CP$ asymmetry. 

This paper is organized as follows. Section~\ref{sec:Formalism} provides a brief outline of the formalism used to calculate the amplitude, the differential mass distribution, and the $CP$ asymmetry. The results are then presented and examined in Section~\ref{sec:Results}. Finally, we conclude with a summary and discussion of the main findings in Section~\ref{sec:Conclusion}.

\section{Formalism}
\label{sec:Formalism} 

We begin by describing the formalism used to compute the total amplitude for the decay $B^- \to K^+ K^-  \pi^-$. This amplitude is given by the sum of the tree-level and triangle-loop contributions at the hadron level and serves as the input for evaluating the invariant mass distribution and the associated $A_{CP}$. The corresponding amplitude for the $B^+$ decay can be obtained analogously; therefore, for conciseness, we discuss only the $B^-$ case in detail.

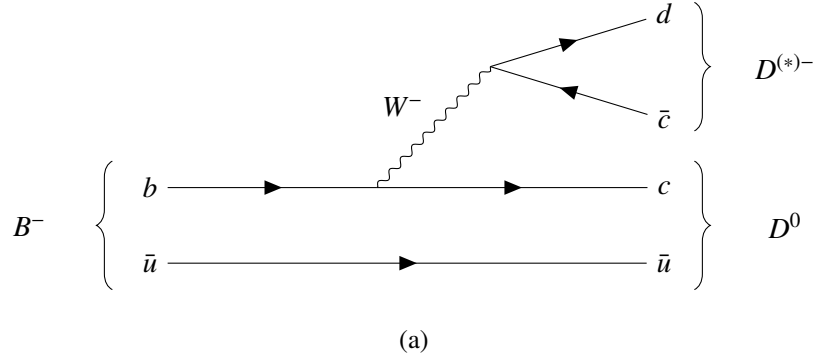
\begin{figure}[!htbp]
\subfigure[]{
\begin{tikzpicture}
\begin{feynman}
\vertex (b0) at (0,1) {\(b\)};
\vertex (u0) at (0,0) {\(\bar u\)};
\vertex (v1) at (3,1);
\vertex (u1) at (6.8,1) {\( c \)};
\vertex (u2) at (6.8,0) {\(\bar u \)};
\vertex (w) at (4.5,2.6);
\vertex (d) at (6.8,3.3) {\(d\)};
\vertex (ubar) at (6.8,1.9) {\(\bar c\)};
\diagram*{
(b0) -- [fermion] (v1) -- [fermion] (u1),
(u0) -- [fermion] (u2),
(v1) -- [boson, edge label=\(W^-\)] (w),
(w) -- [fermion] (d),
(ubar) -- [fermion] (w),
};
\draw[decorate,decoration={brace,mirror,amplitude=6pt}]
(-0.5,1.35) -- (-0.5,-0.35)
node[midway,xshift=-1.1cm] {\(B^-\)};
\draw[decorate,decoration={brace,amplitude=6pt}]
(7.2,3.45) -- (7.2,1.75)
node[midway,xshift=1.2cm] {\( D^{(\ast)-}\)};
\draw[decorate,decoration={brace,amplitude=6pt}]
(7.2,1.35) -- (7.2,-0.35)
node[midway,xshift=1.2cm] {\( D^0 \)};
\end{feynman}
\end{tikzpicture}
}
\caption{
    Quark-level diagram contributing to the high-mass region of the $B^-\to K^+K^- \pi^- $ process. The $W^{-}$ external-emission  $b\to c $ produces a $\bar D^* D $ pair, which subsequently rescatters to reach the final state.}
\label{d1}
\end{figure}

The two most relevant weak decays are the $W$ external emission illustrated in Figs.~\ref{fig1b} and~\ref{d1}, i.e. neglecting penguin and $W$ internal emission, that we refer as tree and rescattering diagrams. For the tree-level contribution, the $ b$ quark decays into a $u$ quark via the external emission of a $W^-$ boson, which subsequently decays into a $d\bar u$ pair to form a $\pi^-$ meson. 
The $\bar u$ quark from the $B^-$ acts as a spectator and, together with the $u$ quark produced in the external $W^-$ emission, hadronizes with an $s\bar s$ pair generated from the vacuum, giving rise to the $K^+ K^-$ pair in the final state. This channel will give source to $K \bar K$ isospin zero resonances as observed in the recently amplitude analysis~\cite{LHCb:2019xmb}.
The rescattering mechanism arises from the $\bar b$ quark decaying into a $c$ quark through the external emission of a $W^-$ boson, which decays into a $d\bar c$ pair to form a $D^{(\ast)-}$ meson. In this case, the spectator $\bar u$ quark from the $B^-$ combines with the $c$ quark from the $W^-$ emission to form a $D^0$ meson. These charmed mesons will be the intermediate particles participating in the triangle loop to yield the final state. 


\subsection{Tree-level Amplitude}


The tree level diagram of Fig.~\ref{fig1b} contains a $b\to u$ weak transition carrying a CKM weak phase. Once we are interested on energies above $c \bar c$ threshold, we parametrize the tree-level contribution as a constant amplitude, given that only resonance tails will be present to interfere with the rescattering amplitude. 
Consequently, this term accounts for the kinematic behavior of the decay and reproduces the phase-space distribution, while the nontrivial dynamical features are encoded in the loop contribution. The hadronic loop term, which contains the strong phase, will interfere with the non-resonant contribution that carries the weak phase, thereby providing the necessary weak and strong-phase difference for CP violation. 


Accordingly, the amplitude describing the contribution of the tree-level mechanism 
is given by $ \mathcal{M}^{(\text{tree})} = C_{0}, $ where $C_0$ is a strength parameter fixed so that the resulting branching fraction is smaller than the reaction $B^-\to K^+K^- \pi^- $, yet of the same order, namely $\mathcal{B}(B^- \to \pi^- K^+K^-) = 5.3 \times 10^{-3}$~\cite{ParticleDataGroup:2024cfk}.



\subsection{Triangle loop}

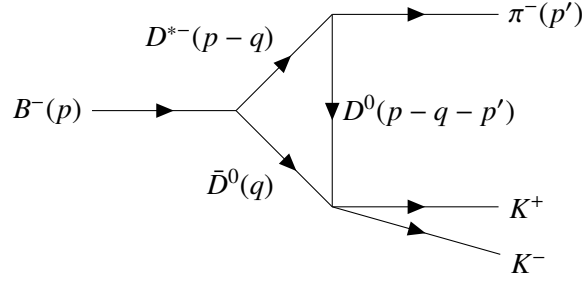
\begin{figure}[!htbp]
\begin{tikzpicture}
\begin{feynman}
  \vertex (B) {$B^-(p)$};
  \vertex [right=2.5cm of B, dot, minimum size=3pt] (v1);

  \vertex [above right=1.8cm of v1, dot, minimum size=3pt] (v2);
  \vertex [below right=1.8cm of v1, dot, minimum size=3pt] (v3);

  \vertex [ right=2.2cm of v2] (pi) {$\pi^-(p')$};
  \vertex [right=2.2cm of v3] (k1) {$K^+$};
  \vertex [below =0.6cm of v3] (k2prime) {};
  \vertex [right =2.6cm of k2prime] (k2) {$K^-$};

  \diagram*{
    (B) -- [fermion] (v1),

    (v1) -- [fermion, edge label=$D^{*-}(p-q)$] (v2),
    (v2) -- [fermion, edge label=$D^0(p-q-p')$] (v3),
    (v3) -- [anti fermion, edge label=$\bar D^0(q)$] (v1),

    (v2) -- [fermion] (pi),
    (v3) -- [fermion] (k1),
    (v3) -- [fermion] (k2),
  };
\end{feynman}
\end{tikzpicture}
\caption{Triangle Feynman diagram for the decay  $B^-\to K^+K^- \pi^- $ via the following charmed meson loop: $B^- \to  D^{\ast+} (\to \pi^{ -} D^{0} ) \bar D^{0} \to \pi^{ -} (   D^{0}  \bar D^{0} ) \to \pi^- K^+K^- $. The respective momenta of the particles are in parentheses. }
\label{d2}
\end{figure}

For the so called rescattering contribution,  the hadron-level realization of the process  begins with the weak vertex, depicted in Fig.~\ref{d1}, and proceeds through the triangle mechanism illustrated in Fig.~\ref{d2} to yield the final state. The calculation of the corresponding amplitude will be carried out within the effective Lagrangian framework, as discussed in the following.

The primary weak decay $B^- \to D^{\ast -} \bar D^{0}$ is treated within an effective formalism. A detailed justification of this methodology, which draws inspiration from the factorization approach, can be found in Refs.~\cite{Ali:1998eb,Wang:2022xga,Abreu:2025mhl}. The adopted effective approach captures the essential dynamics of the weak process, providing a simplified and tractable framework for this exploratory analysis of the rescattering effects.
Concretely, the production amplitude for the decay mode $B^- (p) \to D^{\ast -} (p_1) \bar D^{0} (p_2)$ is treated as follows:
\begin{eqnarray}
 t _{B^- D^{\ast -} \bar D^{0} }  & = &  C \ (p+p_2)_{\mu} \ \epsilon_{D^{*-}}^\mu (p_1) , 
  \label{amplBDDS}
\end{eqnarray}
where $C$ is the effective coupling constant to be determined, and $\epsilon^\mu(p_i)$ denotes the polarization vector of the vector meson $D^{*-}$. 
The magnitude of $C$ can be estimated using the expression for the decay rate:
\begin{eqnarray} 
\Gamma( B^- \to D^{\ast -} \bar D^{0} ) = \frac{1}{8\pi}\frac{|\vec{p_1}|}{ m^2_{B}} \sum_{\text{Pol}} |   t _{B^- D^{\ast -} \bar D^{0} } |^2 ,
  \label{decayrate1}
\end{eqnarray}
where $|\vec{p_1}| = \lambda ^{1/2} \left(m^2_{B^-}, m^2_{D^{\ast-}}, m^2_{\bar D^{0}} \right) / (2 m_{B^-})$ is the magnitude of the three-momentum of the $D^{*-}$ meson in the $B^-$ rest frame; $\lambda (a,b,c) = a^2 + b^2 + c^2 - 2ab - 2ac - 2bc$ is the Källén function; and $\sum_{\text{Pol}}$ denotes the sum over  final state polarization. 

Replacing the amplitude given in Eq.~(\ref{amplBDDS}) into Eq.~(\ref{decayrate1}) yields:
\begin{eqnarray}
 \frac{ C^2 }{\Gamma_ {B}} = \frac{2 \pi \ m^2_{D^{\ast-} }}{|\vec{p_1}|^3} \times  
 \mathcal{B}( B^- \to D^{\ast -} \bar D^{0} ), 
  \label{couplBDSD}
\end{eqnarray}
where $\mathcal{B}( B^- \to D^{\ast -} \bar D^{0} ) = \Gamma( B^- \to D^{\ast -} \bar D^{0} ) / \Gamma_{B}$ is the branching fraction of the corresponding reaction, with $\Gamma_{B}$ being the total decay width of the $B^-$ meson. Experimental data for this branching fraction are available in Ref.~\cite{ParticleDataGroup:2024cfk}. 
Fitting to the central value of the branching ratio yields the coupling constant $C$:
\begin{align}
    \frac{C^2}{\Gamma_B} = 0.00537 \quad\text{MeV}^{-1} . 
\label{const_C}
\end{align}
To describe the $D^{*-} \pi^- D^0$ vertex, we employ an effective Lagrangian based on a pseudoscalar-pseudoscalar-vector (PPV) interaction. This approach follows a generalized $SU(4)$ version of the hidden gauge formalism~\cite{Bando:1984ej,Bando:1987br,Harada:2003jx,Meissner:1987ge}, in which vector mesons are interpreted as dynamical gauge bosons of the hidden $SU(4)_V$ local symmetry within the $U(4)_L \times U(4)_R / U(4)_V$ nonlinear sigma model~\cite{MartinezTorres:2014son}. 

Then, considering only the terms relevant to our analysis, the effective Lagrangian for the $D^{*-} \pi^- D^0$ vertex is given by:
\begin{align}
\mathcal{L}_{D^{*-} \pi^- D^0} = -ig \left( \pi^- \partial_\mu \bar D^0 - \bar D^0 \partial_\mu \pi^- \right) D^{*- \mu},
\label{eq3}
\end{align}
from which the corresponding amplitude takes the form:
\begin{align}
t_{D^{*-} \pi^- D^0} = i g \left( 2p' - p + q \right)_\nu \epsilon^{* \mu}(k).
\label{eq4}
\end{align}
Here the coupling constant $g$ is given by $g = m_V / (2f_\pi)$, with $m_V = 800$ MeV and $f_\pi = 93 $ MeV.

Finally, the remaining ingredient required for the triangle-loop amplitude is the matrix element describing the $D^{0} \bar D^{0} \to K^+ K^-$ transition. This can be obtained within the coupled-channel Bethe-Salpeter formalism, based on the unitarized amplitude $T = [1 - VG]^{-1} V$, where $G$ is the loop function for two intermediate mesons in a given channel, and $V$ is the $S$-wave projected amplitude for the process. This approach naturally allows us to incorporate into our model the contribution from the pole associated with the state $X(3700)$, with $I^{G}(J^{PC}) = 0^{+}(0^{++})$, identified in Ref.~\cite{Xiao:2012iq} by solving the Bethe-Salpeter equation for a potential that includes all possible $PP \to PP$ channels within the local hidden gauge approach with $SU(4)$ symmetry.

However, similarly to Ref.~\cite{Brandao:2023vyg}, and given that the present study is exploratory rather than a high-precision analysis, we benefit from the results of Ref.~\cite{Xiao:2012iq} and describe the $D^{0} \bar D^{0} \to K^+ K^-$ transition amplitude using a Breit–Wigner formula based on the pole information and its couplings. Explicitly, we employ 
\begin{align}
    t_{D^{0} \bar D^{0} , K^+ K^-}(m_{K^+ K^-}) = \frac{g_{D^0\bar D^0} \ g_{K^+K^-}}{m_{K^+ K^-}^2 - m^2_{X} + i \ m_{X} \ \Gamma_{X}},
    \label{eq5}
\end{align}
where $m_{K^+ K^-}$ is the invariant mass of the $K^+ K^-$ system; and the relevant quantities are $g_{D^0\bar D^0} = ( 5.962 + i \ 1.695)$ GeV, $g_{K^+K^-} = (5\times 10^{-3} + i\ 2.2\times 10^{-4})$ GeV, $m_{X} = 3.722 $ GeV, and $ \Gamma_{X} = 36 $ MeV, as taken from Ref.~\cite{Xiao:2012iq}.

Then, with the couplings defined above, the amplitude for the processes depicted in Fig.~\ref{d2} can be written as
\begin{align}
 \mathcal{M} ^{(R)} =  g \ C \int d^4q \ \epsilon^{*\lambda\mu}(p+q)_\mu \ (2p'-p+q)_\nu \ \epsilon^{*\mu} (k)  \ t_{D^{0}  \bar D^{0} , K^+ K^-}(m_{K^+ K^-}),\nonumber\\
\times\frac{1}{(p-q)^2 - m_{D^{*-}} +i\epsilon}\times\frac{1}{(p-q - p')^2 - m_{\bar {D^0}} +i\epsilon}\times \frac{1}{q^2 - m_{{D^0}} +i\epsilon}. 
\label{eq6}
\end{align}

The amplitude $\mathcal{M} ^{(R)}$ in Eq.~(\ref{eq6}) is evaluated according to the following strategy (omitted for the sake of conciseness): (i) the $B^-$ meson rest frame is adopted, i.e. $p^\mu = (m_B,\vec 0)$; (ii) the analytical integration over $q_0$ is performed, picking up the three poles in the contour below the real axis~\cite{Brandao:2023vyg}: $\xi_1 =  \omega_{{D^0}}(\vec q) - i\epsilon, \xi_4 = m_{B^-} + \omega_{{D^{*-}}}(\vec q) - i\epsilon$ and $\xi_6 = (m_{B^-}-p'^0) + \omega_{\bar D^0}(\vec q,\vec p') - i\epsilon$, where:
\begin{align}
\omega_{D}(\vec q)        &\equiv \sqrt{|\vec q|^2 + m_{D^0}^2}, \nonumber\\
\omega_{D^{*-}}(\vec q)      &\equiv \sqrt{|\vec q|^2 + m_{D^{*-}}^2}, \nonumber\\
\omega_{\bar D^0}(\vec q,\vec p') &\equiv \sqrt{|\vec q + \vec p'|^2 + m_{\bar D^0}^2}. \nonumber
\end{align}

\subsection{Total amplitude and CP asymmetry}


Once more, we emphasize that our focus is on the higher-energy region. In this regard, we assume that the low-energy resonances appearing from the $K\pi$ and $K \bar K$ interactions give small tail contributions, which will be roughly incorporated into the tree-level contribution. Thus, we consider only the projected amplitude on the $K\bar{K}$ invariant mass.
Then, the dynamical structure and the contributions of the decay have been determined, the total amplitude can be written as
\begin{align}
 \mathcal{M} (m_{ K^+ K^-}) = e^{\pm i\gamma_{\text{w}}} \ \mathcal{M}^{(\text{tree})} (m_{ K^+ K^-}) + \mathcal{M} ^{(R)}(m_{ K^+ K^-}),  
\label{eq7}
\end{align}
where $\gamma_{\text{w}}$ denotes the weak phase, which is proportional to the $V_{ub}$ term in the Wolfenstein parametrization; the $\pm$ sign corresponds to the $B^{\mp}$ decay mode.

Since our interest also includes the CP asymmetry, it is convenient to introduce a strong phase in the tree-level contribution through the replacement $\mathcal{M}^{(\text{tree})}  \to e^{i\theta} \, \mathcal{M}^{(\text{tree})}  $, as done in Ref.~\cite{Bediaga:2020ztp}. 
In this work we fix the weak phase at $\gamma_{\mathrm{w}} = 70^\circ$~\cite{ParticleDataGroup:2024cfk}, while the strong phase $ \theta $ is treated as a free parameter. As will be shown, this phase, along with the remaining model inputs, affects the invariant-mass distribution.

With these considerations, the CP asymmetry arising from the interference between amplitudes carrying strong and weak phases depends on $\gamma$ and is estimated through
\begin{align}
    A_{CP}(m_{ K^+ K^-}) 
= \frac{|{\cal M}(m_{ K^+ K^-})|^2 - |\overline{\cal M}(m_{ K^+ K^-})|^2}
       {|{\cal M}(m_{ K^+ K^-})|^2 + |\overline{\cal M}(m_{ K^+ K^-})|^2}.
    \label{eq10}
\end{align}
This expression allows us to examine whether the model can reproduce the main features observed in the data~\cite{LHCb:2022fpg}, namely a CP asymmetry that remains mostly negative in the region of interest where the $X(3700)$ emerges and the $D\bar D$ channel opens.

\subsection{Invariant mass distribution}

To obtain the differential mass distribution, we use the usual mass distribution formula,
\begin{align}
    \frac{d\Gamma}{ds_{12}ds_{23}} = \frac{1}{(2\pi^3)}\frac{1}{32m_{B^-}^3}|\mathcal{M}|^2 . 
    \label{eq9}
\end{align}
In the present case, we denote particles $1$, $2$, and $3$ as the mesons $K^+$, $K^-$, and $\pi^-$, respectively. In this sense, the squared invariant masses are defined as $s_{12} \equiv m_{12}^2 \equiv m_{K^+K^-}^2$ and $s_{23} \equiv m_{23}^2 \equiv m_{K^-\pi^-}^2$. Consequently, the mass distribution of the $K^+K^-$ system is obtained by performing the following integration:
\begin{equation}
\frac{d\Gamma}{dm_{K^+K^-}} \equiv \frac{d\Gamma}{d m_{12}} = \frac{1}{(2\pi)^3}\frac{m_{12}}{8 m_{B^-}^3}\ \int\limits_{\left(m_{23}\right)_{\text{min}}}^{\left(m_{23}\right)_{\text{max}}} dm_{23} \  m_{23}  \sum_{\text{Pol}} \left\vert \mathcal{M} \right\vert ^{2}  ,
\label{mass_distr1}
\end{equation}
where 
\begin{align}
\left(m_{23}\right)_{\text{min}}
&=
\left[ \left(E_{K^{+}}^{*} + E_{\pi^{-}}^{*}\right)^2
-
\left(
\sqrt{E_{K^{+}}^{*2} - m_{K}^{2}}
+
\sqrt{E_{\pi^{-}}^{*2} - m_{\pi}^{2}}
\right)^2 \right]^{\frac{1}{2}}
\nonumber\\
\left(m_{23}\right)_{\text{max}}
&=
\left[ \left(E_{K^{+}}^{*} + E_{\pi^{-}}^{*}\right)^2
-
\left(
\sqrt{E_{K^{+}}^{*2} - m_{K}^{2}}
-
\sqrt{E_{\pi^{-}}^{*2} - m_{\pi}^{2}}
\right)^2 \right]^{\frac{1}{2}}
,
\label{mass_distr2}
\end{align}
and
\begin{align}
E_{K^{+}}^{*}
&=
\frac{1}{2}\sqrt{s_{23}}\nonumber\\
E_{\pi^{-}}^{*}
&=
\frac{1}{2\sqrt{s_{23}}}
\left(
m_{B}^{2}
-
s_{23}
-
m_{\pi}^{2}
\right).
\label{mass_distr3}
\end{align}
The variable $m_{12}$ is defined in the range $m_{12} \in [(m_{K^+} + m_{K^-}), (m_B - m_{\pi^-})]$. An analogous procedure can be applied to obtain the result in terms of the other variable $m_{K^- \pi^-}$.

\section{Results}
\label{sec:Results}

Since our model does not account for low-energy dynamics, we restrict our analysis to the kinematic region of interest where the $X(3700)$ resonance is expected to appear. Guided by the specific kinematic cut shown in Fig.~8 of the LHCb data~\cite{LHCb:2022fpg}, we compute the normalized $K^+K^-$ differential mass distribution $\frac{1}{\Gamma_B}\frac{d\Gamma}{dm_{K^+K^-}}$ for the $B^{\mp} \to K^+K^- \pi^{\mp}$ decays. Specifically, we integrate the three-body phase space over the $K^- \pi^{\mp}$ invariant mass squared in the range $3~\text{GeV}^2 \le m_{K^- \pi^{\mp}}^2 \le 16~\text{GeV}^2$ and then examine the resulting $m_{K^+K^-}$ distribution in the interval $3.4~\text{GeV} \le m_{K^+K^-} \le 4~\text{GeV}$, where the $X(3700)$ structure may leave a trace.

\begin{figure}[!htbp]
    \centering
    \includegraphics[width=0.5\linewidth]{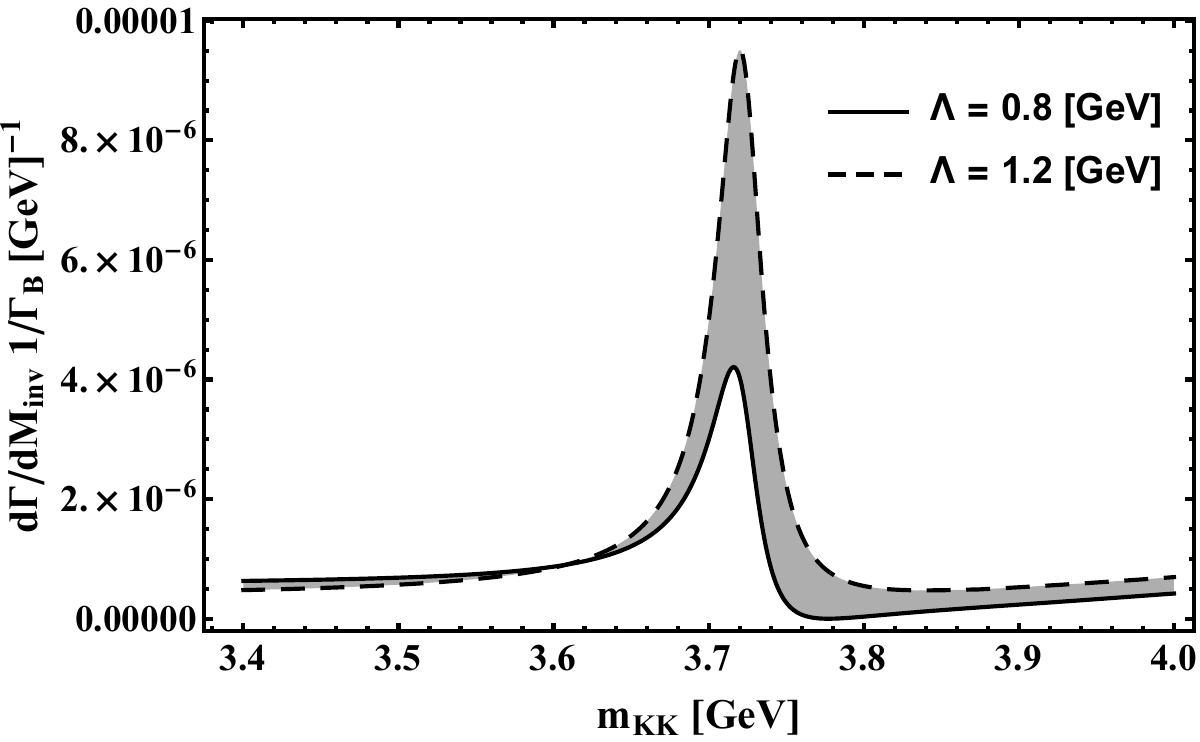}
    \caption{The normalized $K^+K^-$ invariant mass distribution $\frac{1}{\Gamma_B}\frac{d\Gamma}{dm_{K^+K^-}}$ for $B^- \to \pi^- K^+K^-$. The band, which spans the range $0.8 \leq \Lambda \leq 1.2$ GeV, illustrates the theoretical uncertainties of our model. The remaining parameters are fixed at $C_0 = 1$ and $\theta = 0$. } 
    \label{fig:IMDSET1}
\end{figure}

\begin{figure}[!htbp]
    \centering
    \includegraphics[width=0.5\linewidth]{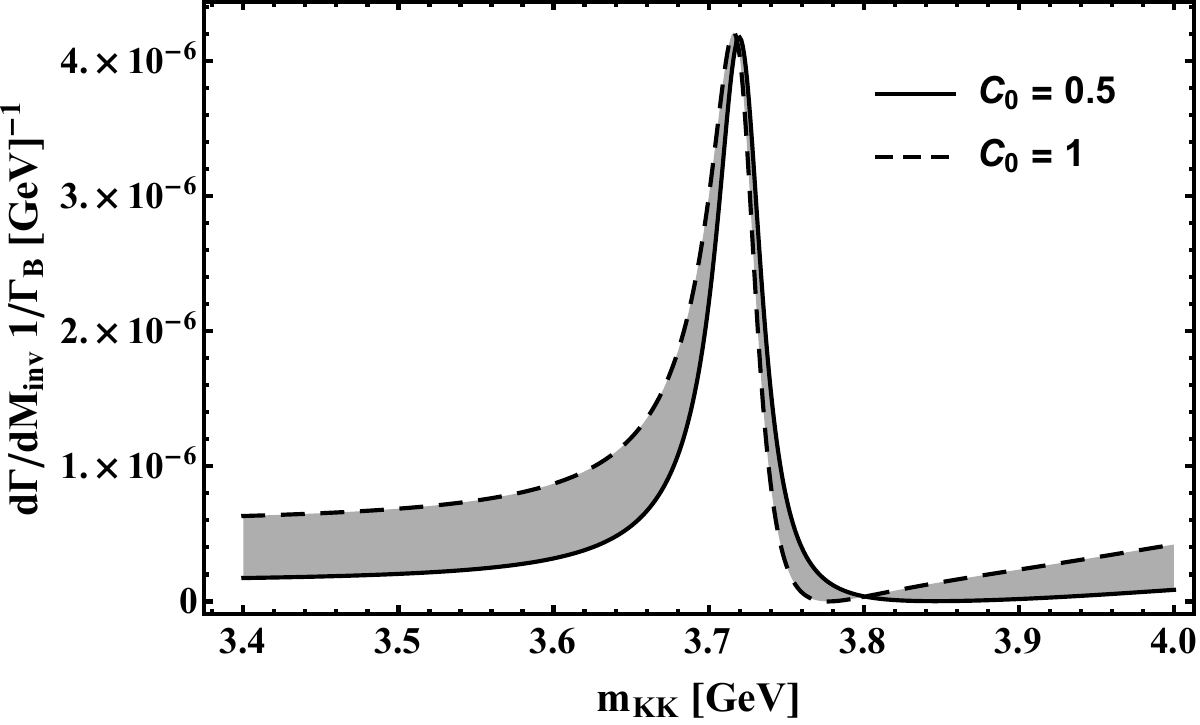}
    \caption{As in Fig.~\ref{fig:IMDSET1}, but with $C_0$ varied in $0.5 \leq C_0 \leq 1.0$. The remaining parameters are fixed at  $\theta = 0$ and $\Lambda = 0.8$ GeV. 
}
    \label{fig:IMDSET3}
\end{figure}


\begin{figure}[!htbp]
    \centering
    \includegraphics[width=0.48\linewidth]{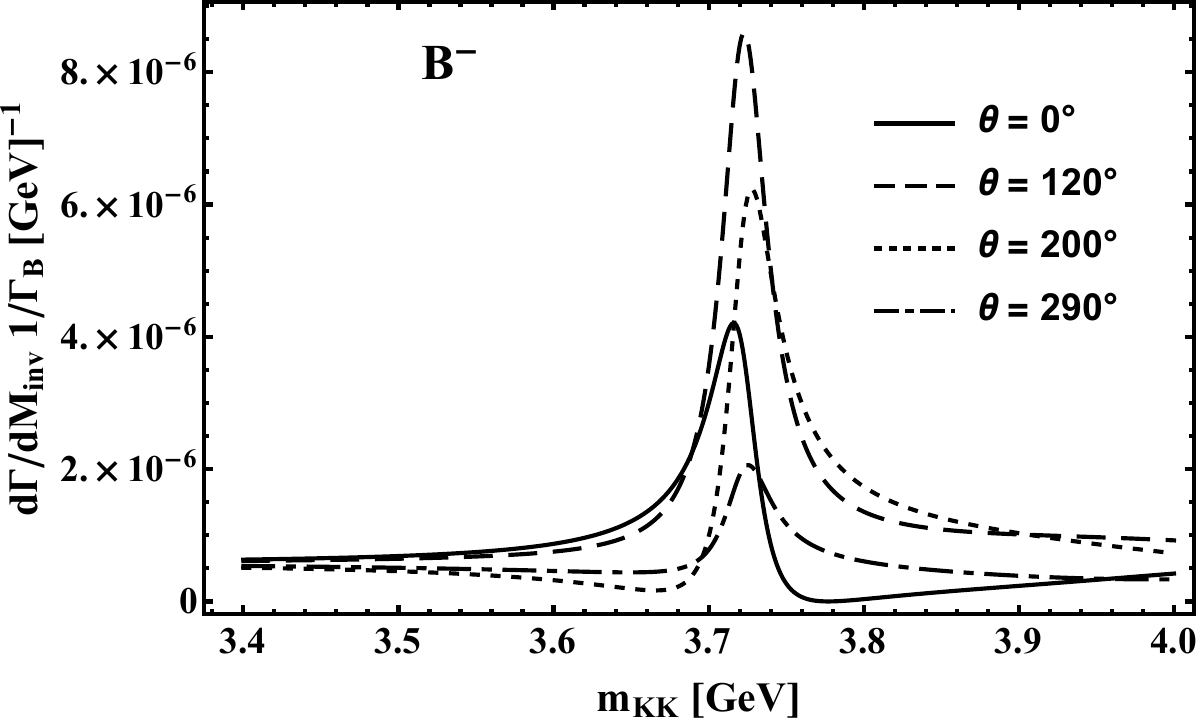}
    \includegraphics[width=0.48\linewidth]{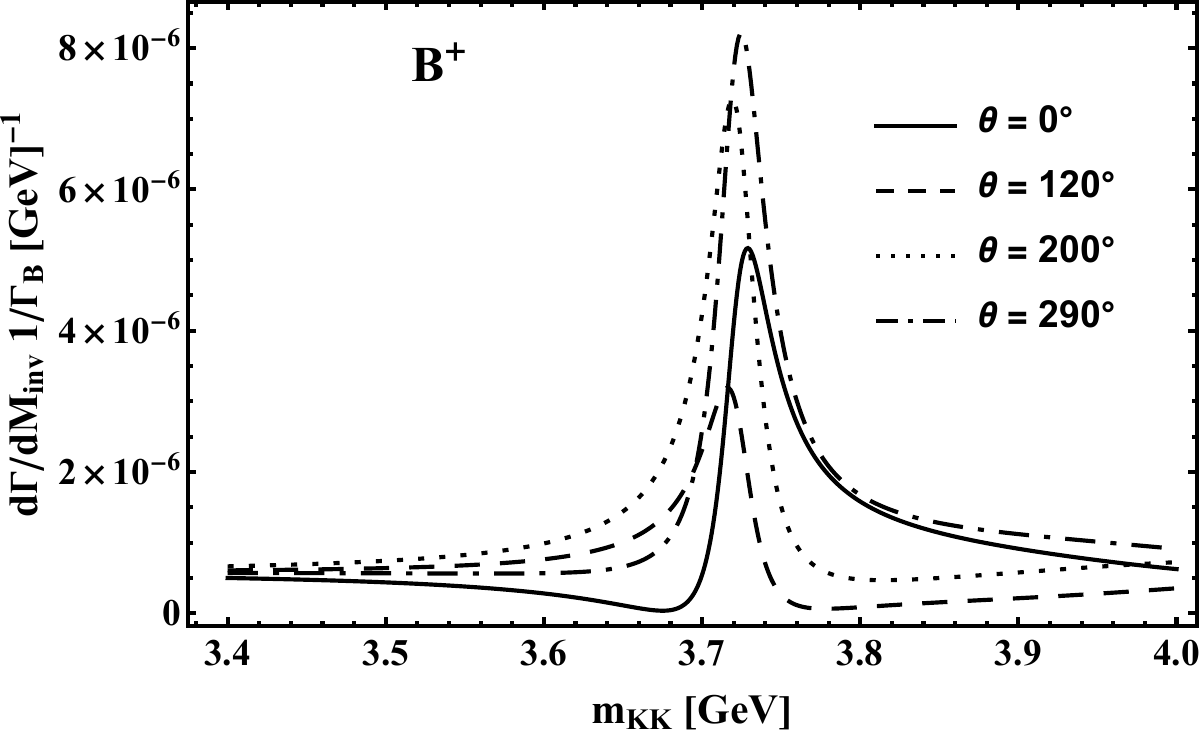 }
    \caption{As in Fig.~\ref{fig:IMDSET1}, but with the strong phase $\theta$ in the tree amplitude, within the range $0^\circ \leq \theta \leq 200^\circ$. Left (right) panel shows the reaction $B^- \to K^+K^-\pi^-$ $(B^+ \to K^+K^-\pi^+)$. The remaining parameters are fixed at $C_0 = 1$ and $\Lambda = 0.8$ GeV. }
    \label{fig:IMDSET2}
\end{figure}

We first present the main features of our model, analyzing how the invariant mass distribution depends on the relevant parameters. Fig.~\ref{fig:IMDSET1} shows the $K^+K^-$ normalized mass distribution for the decay $B^- \to \pi^- K^+K^-$, obtained by inserting the amplitude from Eq.~\eqref{eq7} into Eq.~\eqref{mass_distr1}. The band corresponds to curves for different values of the cutoff parameter $\Lambda$ in the range $0.8 \leq \Lambda \leq 1.2$ GeV, encoding the theoretical uncertainties of our model. A pronounced peak is observed around $3.72$ GeV, near the $D\bar{D}$ threshold, indicating the presence of the $X(3700)$ structure arising from the rescattering contribution. The magnitude of the peak exhibits a clear dependence on $\Lambda$, suggesting that further investigation of this channel may provide valuable insights into the nature of this exotic hadron.

Fig.~\ref{fig:IMDSET3} illustrates the effect of varying the magnitude $C_0$ of the tree-level contribution. For the chosen set of parameters, the curve does not change significantly in the peak region; the variation is most pronounced in the tails below and above the peak. This indicates that the resonance structure itself is robust against changes in the tree-level magnitude, while the background shape is more sensitive to this parameter.

The dependence on the strong phase $\theta$ of the mass distributions for the $B^{\mp} \to K^+K^- \pi^{\mp}$ decays is shown in Fig.~\ref{fig:IMDSET2}. As expected, setting $\theta  \pm \gamma_w = 0$ yields identical curves for both decay processes. However, unlike the previous case, varying $\theta$ leads to a noticeable change in the magnitude of the peak corresponding to the $X(3700)$ generated by the hadronic loop. More importantly, as $\theta$ varies, the mass distributions for $B^{-}$ and $B^{+}$ decays exhibit distinct patterns. This sensitivity to the relative phase between the amplitudes will be crucial for reproducing the experimental mass distributions and $CP$ asymmetry.

Taken together, these figures demonstrate that while the $X(3700)$ signal remains stable under variations of $C_0$, it is particularly sensitive to the cutoff parameter $\Lambda$ and the strong phase $\theta$.  The latter, in particular, offers a key handle for tuning the model to experimental data and will be essential for describing the observed CP asymmetry in the $B^\mp \to K^+K^-\pi^\mp$ decays.

\begin{figure}[!htbp]
    \centering
    \includegraphics[width=0.6\linewidth]{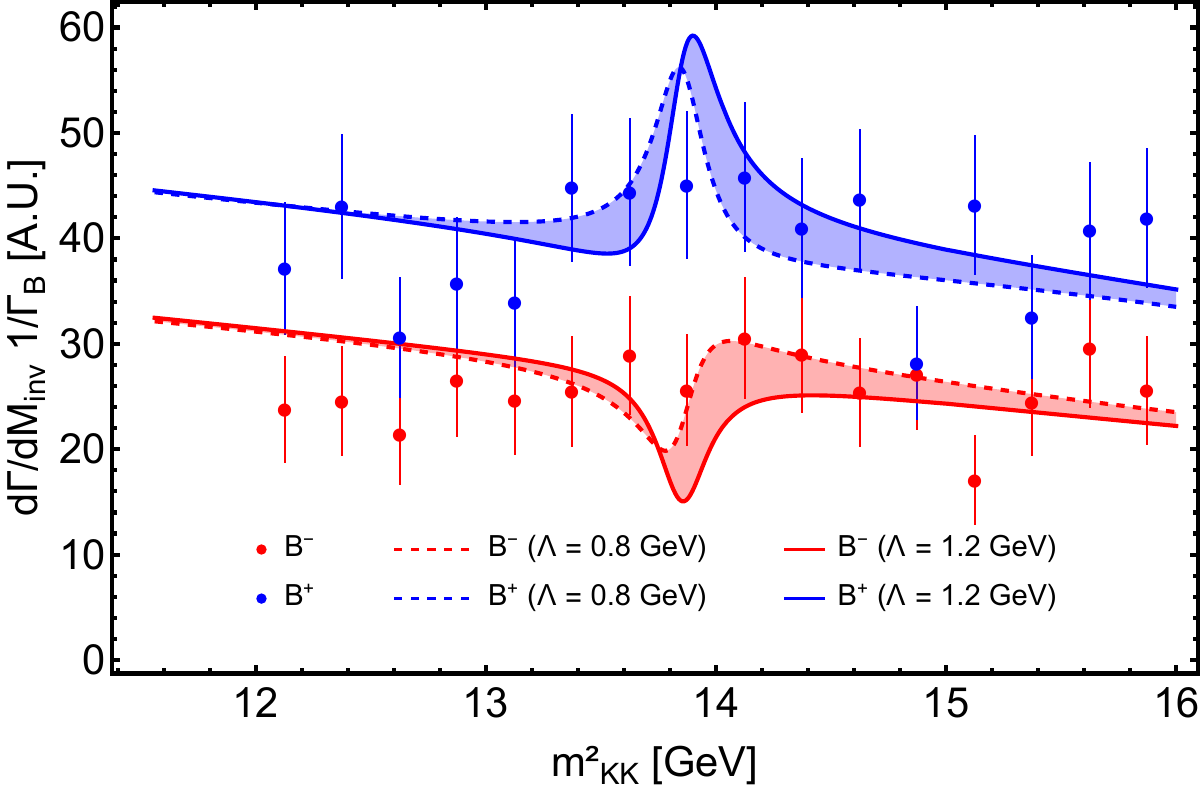}
    \label{fig:IMDSET4a}
    \caption{The $K^+K^-$ differential mass distribution $C_{B^\mp} \times\frac{1}{\Gamma_B}\frac{d\Gamma}{dm_{K^+K^-}}$ for the $B^{\mp} \to K^+K^- \pi^{\mp}$ decays in arbitrary units. The red (blue) curve and data denote the reaction $B^- \to K^+K^-\pi^-$ $(B^+ \to K^+K^-\pi^+)$. The parameters are fixed as $C_{B^{+}} = 5.98 \times 10^{5}$ and $C_{B^{-}} = 4.41 \times 10^{5}$, $C_0 = 13$, and $\theta = 331^{\circ}$. The experimental data are taken from Ref.~\cite{LHCb:2022fpg}. The error band account for varying the triangle integral cutoff over the interval $800 - 1200$ MeV. } 
\label{fig:IMDSET4}
\end{figure}

We now compare our findings with the available data for the differential mass distributions and the $CP$ asymmetry. The parameters are then fixed to achieve a reasonable description of the experimental data from Ref.~\cite{LHCb:2022fpg}.
Here we wish to clarify that our approach was not intended as a least-squares fitting procedure. Instead, we aimed to assess whether the $X(3700)$ signature and $A_{CP}$ in our model could explain the signal observed in the experiments studied. Accordingly, we made natural choices for the parameters of our formalism. The invariant mass distributions for the $B^{\mp}$ reactions are multiplied by a global multiplicative factor (denoted here by $C_{B^\mp}$) to yield the appropriate dimension and magnitude of the data (given in terms of number of events per bin). It provides a reference scale for the overall normalization rather than optimizing the shape of the distribution. The tree-level constant $C_0$, as mentioned before, is fixed so that the resulting branching fraction is of the same order as, but smaller than, the partial one of the considered reactions. Thus, the constant $C_0$ and the strong phase $\theta$ are the free parameters adjusted for best agreement of the theoretical model and the data. The remaining parameter is the cutoff $\Lambda$ in the form factor used to regularize the loop integrals. To account for uncertainties, we varied this cutoff over the interval $\sim 800 - 1200$~MeV.

Fig.~\ref{fig:IMDSET4} shows the $K^+K^-$ invariant mass distribution for the $B^{\mp} \to K^+K^- \pi^{\mp}$ decays. As can be seen, the mass distributions for $B^- \to K^+K^-\pi^-$ and $B^+ \to K^+K^-\pi^+$ differ, with the latter exhibiting larger magnitudes in this specific region, thereby reproducing this feature of the data. 
Another interesting observation is that the results are in reasonable
agreement with the data, particularly in the case of $B^-$, where the
description is compatible with the signature of the $X(3700)$. This suggests that the signature of the $X(3700)$ might be encoded in reactions with rescattering contributions generated by triangle mechanisms involving less relevant channels that have smaller couplings to this structure. In this regard, different channels—such as the one studied here and others discussed in Ref.~\cite{Brandao:2023vyg} that feature stronger couplings to the $X(3700)$—should be analyzed by collaborations like LHCb and Belle to search for traces of this exotic state. Such investigations would provide a more comprehensive understanding of its possible detection and characteristics.

\begin{figure}[!htbp]
    \centering
    \includegraphics[width=0.5\linewidth]{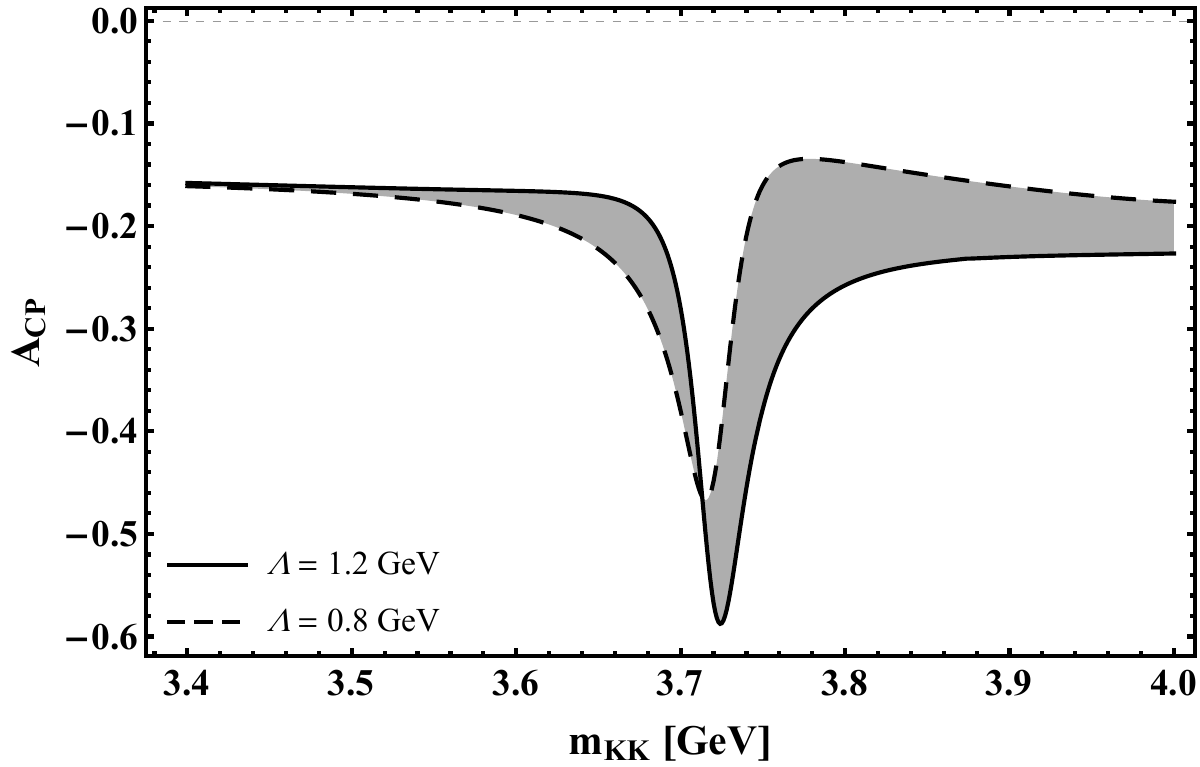}
    \caption{Localized CP asymmetry using the set of parameters as in Fig.~\ref{fig:IMDSET4}. 
    }
    \label{fig:IMDSET5}
\end{figure}

Using these results as inputs, we then estimate the localized $CP$ asymmetry, which is presented in Fig.~\ref{fig:IMDSET5}. 
To compare our findings with the LHCb experimental results~\cite{LHCb:2022fpg}, we extract the data from Fig.~3(c) of Ref.~\cite{LHCb:2022fpg} within the invariant mass interval mentioned above. Since the LHCb data for $A_{CP}$ depend on $m_{K^- \pi^{\mp}}$ across the considered $m_{K^+K^-}$ range, it is more appropriate to integrate the number of events over the $K^- \pi^{\mp}$ invariant mass interval for a fixed value of $m_{K^+K^-}$.
Accordingly, our estimate for $A_{CP}$ is   
\begin{align}
A_{CP}(3.4~\text{GeV} \le m_{K^+K^-} \le 4~\text{GeV}) = [-0.120, -0.142], 
\label{acp1}
\end{align} 
where the error band accounts for varying the triangle integral cutoff over the interval $800 - 1200$ MeV.   
Thus, within the considered region, the localized $CP$ asymmetry assumes a negative value, consistent with the behavior of the curves shown in Fig.~\ref{fig:IMDSET4}. The experimental result from~\cite{LHCb:2022fpg} is
\begin{align}
A_{CP}^{\text{(LHCb)}}(3.4~\text{GeV} \le m_{K^+K^-} \le 4~\text{GeV}) = -0.208 \pm 0.107. 
\label{acp2}
\end{align}
As can be seen, our findings obtained with the appropriate parameter set presents a reasonable agreement with the LHCb data. 

\section{Conclusions}
\label{sec:Conclusion}

We have investigated final-state interaction effects in $B^{\mp}\to K^+K^- \pi^{\mp}$ decays with two main purposes: (i) to study $CP$ asymmetry, and (ii) to search for possible signatures of the $X(3700)$ state. To this end, we have calculated the amplitudes including both tree-level and resonant contributions, with the latter arising from a triangle mechanism $B^{\mp} \to \bar{D}^0 D^{\ast \mp} (\to \pi^{\mp} D^0) \to K^+K^- \pi^{\mp}$, in which the $D\bar D \to K \bar K$ rescattering process carries the signature of the $X(3700)$. The dependence of our results on the relative phase between the tree-level and resonant contributions, as well as on other relevant parameters, has been systematically analyzed.

Our results reveal possible signatures of the $X(3700)$ in the differential mass distributions, suggesting that such signatures may be encoded even in reactions with rescattering contributions generated by triangle mechanisms involving channels that couple weakly to the $X(3700)$, such as the one considered here. When combined with other reactions that feature stronger couplings to this state, these findings would provide a more comprehensive path toward its possible confirmation.

Furthermore, we have estimated a measurable localized $CP$ asymmetry. When compared with recent LHCb data, our findings obtained with an appropriate parameter set show reasonable agreement.

We believe that these results may contribute to a better understanding of the properties of exotic states and the role of $CP$ asymmetry in $B$-meson phenomenology. 

\begin{acknowledgements}

This work was partly supported by the Brazilian agencies CNPq  (Conselho Nacional de Desenvolvimento Cient\'ifico e Tecnol\'ogico) (L.M.A.: Grants No. 400215/2022-5, 308299/2023-0, 402942/2024-8) and CNPq/FAPERJ under the Project INCT-F\'isica Nuclear e Aplica\c{c}\~oes (Contract No. 408419/2024-5).

\end{acknowledgements}


\bibliographystyle{apsrev4-2}
\bibliography{references-ACP}

@article{Gronau_2008,
   title={B decays dominated by  $\omega - \phi$ mixing},
   volume={666},
   ISSN={0370-2693},
   url={http://dx.doi.org/10.1016/j.physletb.2008.07.016},
   DOI={10.1016/j.physletb.2008.07.016},
   number={2},
   journal={Physics Letters B},
   publisher={Elsevier BV},
   author={Gronau, Michael and Rosner, Jonathan L.},
   year={2008},
   month=Aug, pages={185–188} }

@article{Heuser:2025mnk,
    author = "Heuser, L. A. and Reyes-Torrecilla, A. and Hanhart, C. and Kubis, B. and Magalh{\~a}es, P. C. and Mannel, T. and Pel{\'a}ez, J. R.",
    title = "{Understanding Large Localized CP Violation in $B^\pm\to K^\pm\pi^+\pi^-$ Using Dispersive Methods}",
    eprint = "2508.10989",
    archivePrefix = "arXiv",
    primaryClass = "hep-ph",
    doi = "10.1103/l4rs-4znv",
    journal = "Phys. Rev. Lett.",
    volume = "136",
    number = "11",
    pages = "111901",
    year = "2026"
}

@article{Bediaga:2020qxg,
    author = {Bediaga, Ignacio and G{\"o}bel, Carla},
    title = "{Direct $CP$ violation in beauty and charm hadron decays}",
    eprint = "2009.07037",
    archivePrefix = "arXiv",
    primaryClass = "hep-ex",
    doi = "10.1016/j.ppnp.2020.103808",
    journal = "Prog. Part. Nucl. Phys.",
    volume = "114",
    pages = "103808",
    year = "2020"
}

@article{ParticleDataGroup:2024cfk,
    author = "Navas, S. and others",
    collaboration = "Particle Data Group",
    title = "{Review of particle physics}",
    doi = "10.1103/PhysRevD.110.030001",
    journal = "Phys. Rev. D",
    volume = "110",
    number = "3",
    pages = "030001",
    year = "2024"
}

@article{LHCb:2013lcl,
    author = "Aaij, Roel and others",
    collaboration = "LHCb",
    title = "{Measurement of CP violation in the phase space of $B^{\pm} \rightarrow K^{+} K^{-} \pi^{\pm}$ and $B^{\pm} \rightarrow \pi^{+} \pi^{-} \pi^{\pm}$ decays}",
    eprint = "1310.4740",
    archivePrefix = "arXiv",
    primaryClass = "hep-ex",
    reportNumber = "LHCB-PAPER-2013-051, CERN-PH-EP-2013-190",
    doi = "10.1103/PhysRevLett.112.011801",
    journal = "Phys. Rev. Lett.",
    volume = "112",
    number = "1",
    pages = "011801",
    year = "2014"
}

@article{LHCb:2014mir,
    author = "Aaij, Roel and others",
    collaboration = "LHCb",
    title = "{Measurements of $CP$ violation in the three-body phase space of charmless $B^{\pm}$ decays}",
    eprint = "1408.5373",
    archivePrefix = "arXiv",
    primaryClass = "hep-ex",
    reportNumber = "CERN-PH-EP-2014-203, LHCB-PAPER-2014-044",
    doi = "10.1103/PhysRevD.90.112004",
    journal = "Phys. Rev. D",
    volume = "90",
    number = "11",
    pages = "112004",
    year = "2014"
}

@article{LHCb:2019xmb,
    author = "Aaij, Roel and others",
    collaboration = "LHCb",
    title = "{Amplitude analysis of $B^{\pm} \to \pi^{\pm} K^{+} K^{-}$ decays}",
    eprint = "1905.09244",
    archivePrefix = "arXiv",
    primaryClass = "hep-ex",
    reportNumber = "LHCb-PAPER-2018-051, CERN-EP-2019-062",
    doi = "10.1103/PhysRevLett.123.231802",
    journal = "Phys. Rev. Lett.",
    volume = "123",
    number = "23",
    pages = "231802",
    year = "2019"
}

@article{LHCb:2019jta,
    author = "Aaij, Roel and others",
    collaboration = "LHCb",
    title = "{Observation of Several Sources of $CP$ Violation in $B^+ \to \pi^+ \pi^+ \pi^-$ Decays}",
    eprint = "1909.05211",
    archivePrefix = "arXiv",
    primaryClass = "hep-ex",
    reportNumber = "LHCb-PAPER-2019-018, CERN-EP-2019-156",
    doi = "10.1103/PhysRevLett.124.031801",
    journal = "Phys. Rev. Lett.",
    volume = "124",
    number = "3",
    pages = "031801",
    year = "2020"
}

@article{LHCb:2019sus,
    author = "Aaij, Roel and others",
    collaboration = "LHCb",
    title = "{Amplitude analysis of the $B^+ \rightarrow \pi^+\pi^+\pi^-$ decay}",
    eprint = "1909.05212",
    archivePrefix = "arXiv",
    primaryClass = "hep-ex",
    reportNumber = "LHCb-PAPER-2019-017, CERN-EP-2019-157",
    doi = "10.1103/PhysRevD.101.012006",
    journal = "Phys. Rev. D",
    volume = "101",
    number = "1",
    pages = "012006",
    year = "2020"
}

@article{LHCb:2022fpg,
    author = "Aaij, R. and others",
    collaboration = "LHCb",
    title = "{Direct CP violation in charmless three-body decays of B{\ensuremath{\pm}} mesons}",
    eprint = "2206.07622",
    archivePrefix = "arXiv",
    primaryClass = "hep-ex",
    reportNumber = "LHCb-PAPER-2021-049, CERN-EP-2022-100",
    doi = "10.1103/PhysRevD.108.012008",
    journal = "Phys. Rev. D",
    volume = "108",
    number = "1",
    pages = "012008",
    year = "2023"
}

@article{Bediaga:2017axw,
    author = "Bediaga, Ignacio and Frederico, Tobias and Magalh{\~a}es, Patricia C",
    title = "{Charm Penguin in $B^\pm \to K^\pm K^+ K^-$: partonic and hadronic loops}",
    eprint = "1712.01230",
    archivePrefix = "arXiv",
    primaryClass = "hep-ph",
    doi = "10.1016/j.physletb.2018.02.062",
    journal = "Phys. Lett. B",
    volume = "780",
    pages = "357--362",
    year = "2018"
}

@article{Bediaga:2020ztp,
    author = "Bediaga, I. and Frederico, T. and Magalhaes, P. C.",
    title = "{CP asymmetry from hadronic charm rescattering in $B^\pm \to \pi^-\pi^+\pi^\pm$ decays at the high mass region}",
    eprint = "2003.10019",
    archivePrefix = "arXiv",
    primaryClass = "hep-ph",
    doi = "10.1016/j.physletb.2020.135490",
    journal = "Phys. Lett. B",
    volume = "806",
    pages = "135490",
    year = "2020"
}

@article{magalhaes2023,
    author = "Magalhães, P. C. and Bediaga, I. and Frederico, T.",
    title = "{Triangle Singularity in $B^+ \to K^+ \pi^- D_s^+ D^{*0}$ Decay}",
    journal = "Phys. Rev. Lett.",
    volume = "131",
    pages = "051802",
    year = "2023",
    doi = "10.1103/PhysRevLett.131.051802"
}

@article{Gamermann:2006nm,
    author = "Gamermann, D. and Oset, E. and Strottman, D. and Vicente Vacas, M. J.",
    title = "{Dynamically generated open and hidden charm meson systems}",
    eprint = "hep-ph/0612179",
    archivePrefix = "arXiv",
    doi = "10.1103/PhysRevD.76.074016",
    journal = "Phys. Rev. D",
    volume = "76",
    pages = "074016",
    year = "2007"
}

@article{Gamermann:2007mu,
    author = "Gamermann, D. and Oset, E.",
    title = "{Hidden charm dynamically generated resonances and the e+ e- ---\ensuremath{>} J / psi D anti-D, J / psi D anti-D* reactions}",
    eprint = "0712.1758",
    archivePrefix = "arXiv",
    primaryClass = "hep-ph",
    doi = "10.1140/epja/i2007-10580-5",
    journal = "Eur. Phys. J. A",
    volume = "36",
    pages = "189--194",
    year = "2008"
}

@article{Wang:2019evy,
    author = "Wang, En and Liang, Wei-Hong and Oset, Eulogio",
    title = "{Analysis of the $e^+e^- \to J/\psi D \bar D$ reaction close to the threshold concerning claims of a $\chi _{c0}(2P)$ state}",
    eprint = "1902.06461",
    archivePrefix = "arXiv",
    primaryClass = "hep-ph",
    doi = "10.1140/epja/s10050-021-00351-9",
    journal = "Eur. Phys. J. A",
    volume = "57",
    pages = "38",
    year = "2021"
}

@article{Ding:2020dio,
    author = "Ding, Zuo-Ming and Jiang, Han-Yu and He, Jun",
    title = "{Molecular states from $D^{(*)}\bar{D}^{(*)}/B^{(*)}\bar{B}^{(*)}$ and $D^{(*)}D^{(*)}/\bar{B}^{(*)}\bar{B}^{(*)}$ interactions}",
    eprint = "2011.04980",
    archivePrefix = "arXiv",
    primaryClass = "hep-ph",
    doi = "10.1140/epjc/s10052-020-08754-6",
    journal = "Eur. Phys. J. C",
    volume = "80",
    number = "12",
    pages = "1179",
    year = "2020"
}

@article{Xiao:2012iq,
    author = "Xiao, C. W. and Oset, E.",
    title = "{Three methods to detect the predicted $D \bar{D}$ scalar meson X(3700)}",
    journal = "Eur. Phys. J. A",
    volume = "49",
    pages = "52",
    year = "2013",
    doi = "10.1140/epja/i2013-13052-5",
    eprint = "1211.1862",
    archivePrefix = "arXiv",
    primaryClass = "hep-ph"
}

@article{Brandao:2023vyg,
    author = "Brandão, P. C. S. and Song, J. and Abreu, L. M. and Oset, E.",
    title = "{B+ decay to K+ $\eta$ $\eta$ with ($\eta$ $\eta$) from the D$\bar{D}$(3720) bound state}",
    journal = "Phys. Rev. D",
    volume = "108",
    number = "5",
    pages = "054004",
    year = "2023",
    doi = "10.1103/PhysRevD.108.054004",
    eprint = "2307.05351",
    archivePrefix = "arXiv",
    primaryClass = "hep-ph"
}

@article{Dai:2020yfu,
    author = "Dai, Lianrong and Toledo, Genaro and Oset, Eulogio",
    title = "{Searching for a $D {\bar{D}}$ bound state with the $\psi (3770) \rightarrow \gamma D^0 {\bar{D}}^0$ decay}",
    eprint = "2004.05204",
    archivePrefix = "arXiv",
    primaryClass = "hep-ph",
    doi = "10.1140/epjc/s10052-020-8058-8",
    journal = "Eur. Phys. J. C",
    volume = "80",
    number = "6",
    pages = "510",
    year = "2020"
}

@article{Sobrinho:2024tre,
    author = "Sobrinho, F. C. and Abreu, L. M. and Bertulani, C. A. and Navarra, F. S.",
    title = "{Production of meson molecules in ultraperipheral heavy ion collisons}",
    eprint = "2405.02645",
    archivePrefix = "arXiv",
    primaryClass = "hep-ph",
    doi = "10.1103/PhysRevD.110.034037",
    journal = "Phys. Rev. D",
    volume = "110",
    number = "3",
    pages = "034037",
    year = "2024"
}

@article{Abreu:2025mhl,
    author = "Abreu, Luciano M. and Torres-Rincon, Juan M.",
    title = "{Searching for femtoscopic signatures of the $D \bar D (I=0) \ [X(3700)]$ bound state}",
    eprint = "2503.13668",
    archivePrefix = "arXiv",
    primaryClass = "hep-ph",
    doi = "10.1103/PhysRevD.xxx.xxxxxx",
    journal = "Phys. Rev. D",
    volume = "",
    pages = "",
    year = "2025"
}

@article{Wang:2020elp,
    author = "Wang, En and Li, Hong-Shen and Liang, Wei-Hong and Oset, Eulogio",
    title = "{Analysis of the $ \boldsymbol{\gamma\gamma \to D\bar D}$ reaction and the $D\bar{D}$ bound state}",
    eprint = "2010.15431",
    archivePrefix = "arXiv",
    primaryClass = "hep-ph",
    doi = "10.1103/PhysRevD.103.054008",
    journal = "Phys. Rev. D",
    volume = "103",
    number = "5",
    pages = "054008",
    year = "2021"
}

@article{Deineka:2021aeu,
    author = "Deineka, Oleksandra and Danilkin, Igor and Vanderhaeghen, Marc",
    title = "{Dispersive analysis of the \ensuremath{\gamma}\ensuremath{\gamma}\textrightarrow{}DD\textasciimacron{} data and the confirmation of the DD\textasciimacron{} bound state}",
    eprint = "2111.15033",
    archivePrefix = "arXiv",
    primaryClass = "hep-ph",
    doi = "10.1016/j.physletb.2022.136982",
    journal = "Phys. Lett. B",
    volume = "827",
    pages = "136982",
    year = "2022"
}

@article{Belle:2007woe,
    author = "Pakhlov, P. and others",
    collaboration = "Belle",
    title = "{Production of New Charmoniumlike States in e+ e- --\ensuremath{>} J/psi D(*) anti-D(*) at s**(1/2) \textasciitilde{} 10. GeV}",
    eprint = "0708.3812",
    archivePrefix = "arXiv",
    primaryClass = "hep-ex",
    doi = "10.1103/PhysRevLett.100.202001",
    journal = "Phys. Rev. Lett.",
    volume = "100",
    pages = "202001",
    year = "2008"
}

@article{Belle:2017egg,
    author = "Chilikin, K. and others",
    collaboration = "Belle",
    title = "{Observation of an alternative $\chi_{c0}(2P)$ candidate in $e^+ e^- \rightarrow J/\psi D \bar{D}$}",
    eprint = "1704.01872",
    archivePrefix = "arXiv",
    primaryClass = "hep-ex",
    reportNumber = "BELLE-PREPRINT-2017-07, KEK-PREPRINT-2017-3",
    doi = "10.1103/PhysRevD.95.112003",
    journal = "Phys. Rev. D",
    volume = "95",
    pages = "112003",
    year = "2017"
}

@article{BaBar:2010jfn,
    author = "Aubert, B. and others",
    collaboration = "BaBar",
    title = "{Observation of the $\chi_{c2}(2p)$ Meson in the Reaction $\gamma \gamma \to D \bar{D}$ at {BaBar}}",
    eprint = "1002.0281",
    archivePrefix = "arXiv",
    primaryClass = "hep-ex",
    reportNumber = "BABAR-PUB-09-031, SLAC-PUB-13939",
    doi = "10.1103/PhysRevD.81.092003",
    journal = "Phys. Rev. D",
    volume = "81",
    pages = "092003",
    year = "2010"
}

@article{Bando:1984ej,
  author = {Bando, M. and Kugo, T. and Uehara, S. and Yamawaki, K. and Yanagida, T.},
  title = {Is rho Meson a Dynamical Gauge Boson of Hidden Local Symmetry?},
  journal = {Phys. Rev. Lett.},
  volume = {54},
  pages = {1215},
  year = {1985},
  doi = {10.1103/PhysRevLett.54.1215}
}

@article{Bando:1987br,
  author = {Bando, M. and Kugo, T. and Yamawaki, K.},
  title = {Nonlinear Realization and Hidden Local Symmetries},
  journal = {Phys. Rept.},
  volume = {164},
  pages = {217--314},
  year = {1988},
  doi = {10.1016/0370-1573(88)90019-1}
}

@article{Meissner:1987ge,
  author = {Meissner, U. G.},
  title = {Low-Energy Hadron Physics from Effective Chiral Lagrangians with Vector Mesons},
  journal = {Phys. Rept.},
  volume = {161},
  pages = {213},
  year = {1988},
  doi = {10.1016/0370-1573(88)90090-7}
}

@article{Harada:2003jx,
  author = {Harada, M. and Yamawaki, K.},
  title = {Hidden local symmetry at loop: A New perspective of composite gauge boson and chiral phase transition},
  journal = {Phys. Rept.},
  volume = {381},
  pages = {1--233},
  year = {2003},
  doi = {10.1016/S0370-1573(03)00139-X},
  eprint = {hep-ph/0302103},
  archivePrefix = {arXiv},
  primaryClass = {hep-ph}
}

@article{MartinezTorres:2014son,
  author = {Martinez Torres, A. and Khemchandani, K. P. and Navarra, F. S. and Nielsen, M. and Abreu, L. M.},
  title = {On $X(3872)$ production in high energy heavy ion collisions},
  journal = {Phys. Rev. D},
  volume = {90},
  number = {11},
  pages = {114023},
  year = {2014},
  doi = {10.1103/PhysRevD.90.114023},
  eprint = {1405.7583},
  archivePrefix = {arXiv},
  primaryClass = {hep-ph},
  note = {Erratum: Phys. Rev. D \textbf{93}, no.5, 059902 (2016)}
}

@article{Ali:1998eb,
  author = {Ali, A. and Kramer, G. and Lu, C. D.},
  title = {Experimental tests of factorization in charmless nonleptonic two-body B decays},
  journal = {Phys. Rev. D},
  volume = {58},
  pages = {094009},
  year = {1998},
  doi = {10.1103/PhysRevD.58.094009},
  eprint = {hep-ph/9804363},
  archivePrefix = {arXiv},
  primaryClass = {hep-ph}
}

@article{Wang:2022xga,
  author = {Wang, H. N. and Geng, L. S. and Wang, Q. and Xie, J. J.},
  title = {Molecular Nature of X(3872) in B$^{0} \to K^{0}$ X(3872) and B$^{+} \to K^{+}$ X(3872) Decays},
  journal = {Chin. Phys. Lett.},
  volume = {40},
  number = {2},
  pages = {021301},
  year = {2023},
  doi = {10.1088/0256-307X/40/2/021301},
  eprint = {2211.14994},
  archivePrefix = {arXiv},
  primaryClass = {hep-ph}
}

\end{document}